\documentclass[twocolumn,aps,prb,showpacs,superscriptaddress,preprintnumbers,amsmath,amssymb]{revtex4-1}

\usepackage[dvips]{graphicx}
\usepackage{dcolumn}
\usepackage{bm}

\begin{document}

\title{Insulator-Metal Transition in TiGePt: a combined photoelectron spectroscopy,
x-ray absorption spectroscopy, and band structure study}

\author{J.~Gegner}
    \affiliation{II. Physikalisches Institut, Universit\"{a}t zu
    K\"{o}ln, Z\"{u}lpicher Stra{\ss}e 77, 50937 K\"{o}ln, Germany}
\author{M.~Gam\.za}
    \affiliation{ Max-Planck-Institut f\"ur Chemische Physik fester
    Stoffe, N\"othnitzer Stra{\ss}e 40, 01187 Dresden, Germany}
    \affiliation{ Institute of Materials Science, University of Silesia, ul. Bankowa 12, 40-007 Katowice, Poland}
\author{S.-V.~Ackerbauer}
    \affiliation{ Max-Planck-Institut f\"ur Chemische Physik fester
    Stoffe, N\"othnitzer Stra{\ss}e 40, 01187 Dresden, Germany}
\author{N.~Hollmann}
    \affiliation{ Max-Planck-Institut f\"ur Chemische Physik fester
    Stoffe, N\"othnitzer Stra{\ss}e 40, 01187 Dresden, Germany}
\author{Z.~Hu}
    \affiliation{ Max-Planck-Institut f\"ur Chemische Physik fester
    Stoffe, N\"othnitzer Stra{\ss}e 40, 01187 Dresden, Germany}
\author{H.-H.~Hsieh}
\affiliation{Chung Cheng Institute of Technology, National Defense
    University, Taoyuan 335, Taiwan}
\author{H.-J.~Lin}
\affiliation{National Synchrotron Radiation Research Center (NSRRC),
    101 Hsin-Ann Road, Hsinchu 30077, Taiwan}
\author{C.~T.~Chen}
\affiliation{National Synchrotron Radiation Research Center (NSRRC),
    101 Hsin-Ann Road, Hsinchu 30077, Taiwan}
\author{A. Ormeci}
    \affiliation{ Max-Planck-Institut f\"ur Chemische Physik fester
    Stoffe, N\"othnitzer Stra{\ss}e 40, 01187 Dresden, Germany}
\author{A. Leithe-Jasper}
    \affiliation{ Max-Planck-Institut f\"ur Chemische Physik fester
    Stoffe, N\"othnitzer Stra{\ss}e 40, 01187 Dresden, Germany}
\author{H.~Rosner}\email{rosner@cpfs.mpg.de}
    \affiliation{ Max-Planck-Institut f\"ur Chemische Physik fester
    Stoffe, N\"othnitzer Stra{\ss}e 40, 01187 Dresden, Germany}
\author{Yu. Grin}
    \affiliation{ Max-Planck-Institut f\"ur Chemische Physik fester
    Stoffe, N\"othnitzer Stra{\ss}e 40, 01187 Dresden, Germany}
\author{L.H.~Tjeng}
    \affiliation{ Max-Planck-Institut f\"ur Chemische Physik fester
    Stoffe, N\"othnitzer Stra{\ss}e 40, 01187 Dresden, Germany}
\date{\today}

\hyphenation{beam-line photo-emission}

\begin{abstract}
We present a combined experimental and theoretical study of the
electronic structure of the intermetallic compound TiGePt by means
of photoelectron spectroscopy, x-ray absorption spectroscopy and
full potential band structure calculations. It was recently shown
[Ref.\ 1] that TiGePt undergoes a structural phase transition by
heating which is accompanied by a large volume contraction and a
drastic change of physical properties, in particular a large
decrease of the electrical resistivity. The present study revealed
substantial differences in the electronic structure for the two
TiGePt modifications, although they have the same nominal
composition and show similar electron counts for particular
valence band states. Our photoemission experiments and band
structure calculations establish that an insulator-to-metal
transition occurs with an appreciable band broadening and closing
of the band gap.
\end{abstract}

\maketitle

\section{Introduction}

\begin{figure}
  \includegraphics[width=0.4\textwidth]{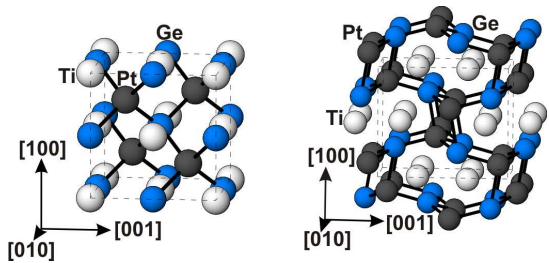}\\
  \caption{(Color online) Crystal structures of the LT phase (cubic MgAgAs-type, left) and the HT phase (orthorhombic TiNiSi-type, right) of TiGePt.\cite{Aackerbauer10} Ti, Ge, Pt atoms are shown as white, blue and grey spheres, respectively.}
\label{structureX}
\end{figure}

The wealth of interesting physical phenomena that can be found for equiatomic intermetallic compounds, like superconductivity, heavy-fermion or Kondo behavior, magnetic ordering, or thermopower, is not only due
to the vast amount of possibilities of combining two or more different elements from the periodic table. The interplay between electronic structure, crystal structure and chemical bonding leads to an additional degree of variability. The ternary equiatomic compound TiGePt is such an example.\cite{Aackerbauer10} This intermetallic adopts two different crystal structures (Fig.~\ref{structureX}). The low temperature (LT) modification of TiGePt forms in the MgAgAs-type  structure\cite{Nowotny41} ("half-Heusler"). Here, Ti and Ge atoms form a sodium chloride type lattice, in which Pt atoms are inserted in half of the tetrahedral voids. By heating above 885 $^{\mathrm{o}}$C, TiGePt transforms via a reconstructive transition into an orthorhombic TiNiSi-type structure\cite{Shoemaker65} with a considerably lowered symmetry. The high temperature (HT) modification of TiGePt can be quenched down to low temperatures.
Its crystal structure can be regarded as a three-dimensional network formed by edge-sharing six-membered puckered rings of Pt and Ge atoms, interlinked along the [100] direction through short Pt-Ge contacts.
In the [010] direction, large eight-membered Pt-Ge rings form channels in which Ti atoms are embedded.

It is remarkable that the volume is reduced by over 10\% in going from the LT to the HT phase.\cite{Aackerbauer10} Furthermore, the occurrence of an insulator-metal transition was suggested based on electrical resistivity measurements.\cite{Aackerbauer10} In the LT phase, TiGePt revealed a semiconducting behavior, while in the HT modification it showed a more metallic temperature dependence with three to four orders of magnitude smaller resistivity values. The structural changes are caused by differences in chemical bonding. Analysis of the atomic interactions within the electron density/electron localizability approach revealed strong differences in atomic interactions between the LT and HT modifications.\cite{Aackerbauer10}

A similar polymorphism has been reported for YbPdSb \cite{Mishra02}, YbAuBi \cite{Merlo90}, GdNiSb \cite{Skolozdra97} and VFeSb \cite{VFeSbT}. All these compounds crystallize in the cubic structure isotypic to MgAgAs at low temperatures. At elevated temperatures, they adopt an AlB$_2$-related crystal structure - being the aristotype of the structure family to which TiNiSi belongs to -  with lower symmetry and larger crystal density.
In the Yb-based systems, the structural transitions are accompanied by changes in valence state of Yb.\cite{Mishra02, Merlo90}
For GdNiSb, an insulator-metal transition has been predicted based on $ab$ $initio$ electronic structure calculations, but not been confirmed experimentally yet.\cite{GdNiSb}
Electrical resistivity measurements for VFeSb suggest a transition from a highly doped semiconductor to a metallic-like conductor at a temperature of 1042 K.\cite{VFeSbT} So far, the change in its electrical properties has not been inspected in detail by means of an electronic structure study.

Here, we report on the electronic structure of TiGePt in both the LT and HT phases. The objective of our study is to establish the relationship between the crystal structure and the electronic properties of TiGePt. To this end we will employ X-ray photoelectron and absorption spectroscopies in combination with full potential electronic band structure calculations.

\section{Methods}
\label{sec:methods}

The samples were
prepared and characterized as described in Ref.~\onlinecite{Aackerbauer10}.
All spectroscopic measurements were carried out at room temperature.
The soft X-ray photoelectron spectroscopy (PES) and absorption
spectroscopy (XAS) experiments were performed at the Dragon
beamline of the NSRRC in Taiwan, using an ultra-high vacuum system
with a pressure in the low 10$^{-10}$ mbar range. For the PES, a
Scienta SES-100 electron energy analyzer was used and the overall
energy resolution was set to 150 meV FWHM at 190 eV photon energy, and
to 350 meV FWHM at 700 eV photon energy. The energy calibration has been done by
using the Fermi cut-off of a polycrystalline Pt metal reference
which was also taken as the zero of the binding energy scale. The
4$f_{7/2}$ core level of the Pt metal was used as an energy
reference.

The XAS spectra at the Ti $L_{2,3}$-edges were
taken in the total electron yield mode with energy resolution of
the photons of 150 meV. A SrTiO$_3$ single crystal was measured
simultaneously as an energy reference for the XAS. Before the
measurements, the polycrystalline TiGePt samples were fractured
\textit{in-situ} to obtain clean surfaces.

The XAS spectra at the Ge $K$--edge were obtained in a transmission
arrangement at the EXAFS beamline C of HASYLAB at DESY, equipped with a Si (111) double
crystal monochromator which yielded an experimental resolution (FWHM) of approximately 3 eV
at the Ge $K$ threshold of about 11100 eV. Powdered materials
were mixed with small amounts of B$_{4}$C and mounted on a  sample holder (1 cm$^{2}$ window) using
paraffin wax. The data were recorded together with powdered Ge as an external reference.

The electronic structure of the two modifications of TiGePt was computed using lattice parameters and atomic positions obtained experimentally at room temperature.\cite{Aackerbauer10}
First--principles band structure calculations were performed using the full--potential local-orbital code FPLO (version 9.01-35)\cite{FPLOKoepernik99} in the fully relativistic mode.
In this method, the four-component Kohn-Sham-Dirac equation containing spin-orbit (SO) coupling to all orders is solved self-consistently.
The Perdew-Wang parametrization\cite{PerdewWang92} of the exchange-correlation potential within the local density approximation (LDA) was employed.
The Brillouin zone was sampled by a well--converged mesh of 27000 $k$-points (30$\times$30$\times$30 mesh, 1368 points in the irreducible wedge of the Brillouin zone) for the cubic LT phase and 10260 $k$-points (20$\times$27$\times$19 mesh, 1540 points in the irreducible wedge of the Brillouin zone) for the orthorhombic HT phase.

To explain the near-edge structures of the Ge $K$ XAS spectra, we carried out band structure calculations by the full potential linearized augmented plane wave (FP--LAPW) method\cite{LAPW} as implemented in the Wien2k\_07 code\cite{Wien2k}. In the scalar--relativistic calculations, exchange-correlation effects were treated within the LDA approximation in the form proposed by Perdew and Wang\cite{PerdewWang92}. Spin-orbit coupling was included in the second variational method using the scalar-relativistic eigenfunctions as basis.\cite{SO}
By comparing the resulting total densities of states (DOS) and band structures with those derived from the fully relativistic calculations using the FPLO code we verified the sufficient accuracy of our FP-LAPW computational results.

The near-edge spectra were calculated according to the formalism described in Refs \onlinecite{XANESt, XANESt1, XANESt2}.
For dipole-allowed transitions, energy dependent matrix elements containing radial transition probabilities were multiplied with the partial DOS. The results were convoluted by the pseudo--Voigt function with a FWHM of 1.5 eV for the Lorentzian and of 2.5 eV  for the Gaussian components, respectively, to mimic the instrumental resolution and the lifetime broadening effects.  Finally, the calculated curves were shifted by 11101.2 eV in order to match the experimental energy scales.

\section{results and discussion}

Fully relativistic electronic structure calculations support the experimental observation that the LT phase is the more stable modification of TiGePt. The calculated difference in the total energy between the two phases is about 0.19 eV per formula unit, which is of the same order as the energy scale of the observed transition temperature of about 1160 K.

\begin{figure}
  \includegraphics[width=0.4\textwidth]{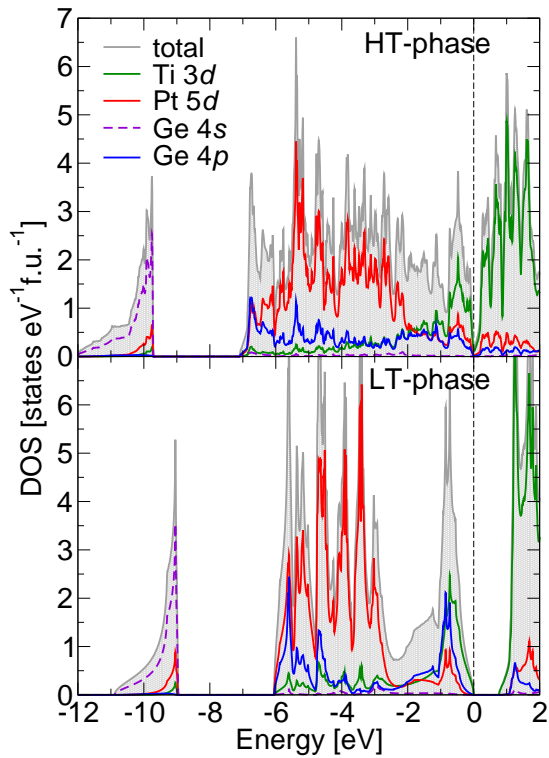}\\
  \caption{(Color online) Total and partial density of states (DOS) from fully relativistic electronic structure calculations of TiGePt: the results for the low-temperature (LT) phase are shown in the bottom panel and the high-temperature (HT) phase in the top panel. The common vertical dashed line indicates the position of the Fermi level.}
\label{LDA}
\end{figure}

Fig.~\ref{LDA} shows the calculated total electronic DOS of the LT phase (bottom panel) and HT phase (top panel) of TiGePt together with the partial DOS of the Pt 5$d$, Ge 4$p$ and 4$s$, and the Ti 3$d$, which are the relevant states composing the valence band. The obtained electron counts for the valence states of the two modifications of TiGePt having the same nominal composition are very similar and amount to about 8.6, 2.7, 1.5, and 2.5 for the Pt 5$d$, Ge 4$p$, Ge 4$s$, and Ti 3$d$ orbital, respectively. Nevertheless, the essential differences in crystal structure and chemical bonding properties between the two phases lead to substantial differences in the DOS as explained below.

For LT-TiGePt there is a band gap of about 0.8~eV, consistent with the semiconducting behavior in the resistivity measurements.\cite{Aackerbauer10} The HT phase, on the other hand, is a metal with a rather low value of the DOS at the Fermi level, $i.e.$ about 0.3 states per eV and formula unit, in line with the results of our thermodynamic and transport study.\cite{Aackerbauer10, Gamta11} For this phase, the DOS above the Fermi level exhibits a pseudogap with a width of about 0.3 eV.

In comparing the valence band of HT-TiGePt with that of the LT phase, one can see immediately that the former has a noticeably larger band width than the latter: 7.2 eV vs 6.0 eV.
The observed band broadening originates from  the altered chemical bonding situation related to the change in local atomic environments, followed by the larger orbital overlap caused by the volume reduction, as argued in Ref. \onlinecite{Aackerbauer10}.
Interestingly, the shallow core Ge 4$s$--like band also broadens accordingly, although it is located far below the Fermi level and is well--separated from the rest of the valence band.
Moreover, this shallow core band is positioned between 9.0 eV and 11 eV binding energy in the LT phase, whereas in HT-TiGePt it is appreciably further away from the Fermi level, namely between 9.7 eV and 12.0 eV binding energy.

The two modifications of TiGePt differ also in the overall shape of the valence band. For the LT phase, the valence band can even be divided into two parts: a~Ti~$3d$--Ge~$4p$ derived band with a sizable admixture of Pt states in the binding energy range from 2 eV to $E_{\mathrm{F}}$ and a broader Pt $5d$ dominated part between 2.5 eV and 6 eV. These features, by contrast, are washed out in the HT phase: one can only recognize that the Pt $5d$ states are more pronounced in the energy region above 2~eV while the Ti $3d$ states contribute more at the lower binding energy part. The Ge $4p$ states are even almost equally distributed over the entire valence band.

One should note that for HT-TiGePt the presented results of the fully relativistic electronic structure calculations are very similar to those obtained recently using the scalar relativistic approach.\cite{Aackerbauer10} In case of  LT-TiGePt, however, the inclusion of the SO coupling has a significant impact on the calculated DOS. It affects the $d$ states of Pt and Ti, the latter ones due to their strong hybridization with the Pt 5$d$ states. Consequently, the width of the valence band is larger than that previously reported and the calculated band gap is smaller by about 0.15 eV.

\begin{figure}
  \includegraphics[width=0.4\textwidth]{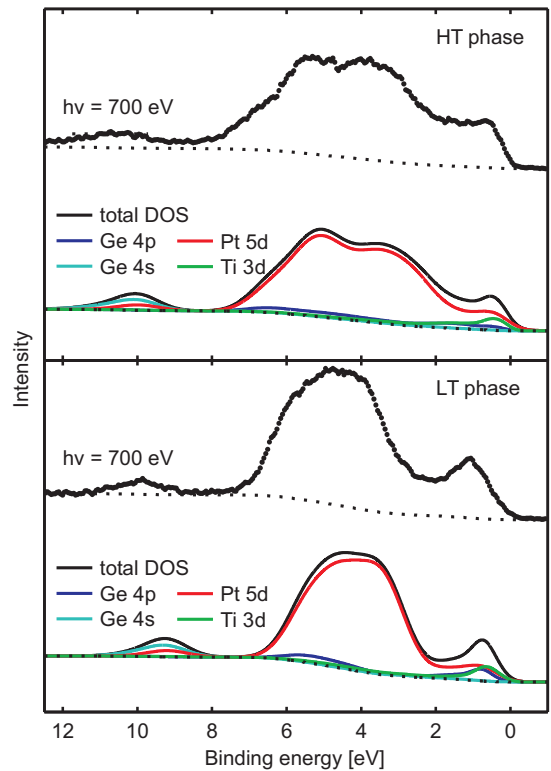}\\
  \caption{(Color online) Valence band spectra of the low-temperature (LT) phase (bottom panel) and the high-temperature (HT) phase of TiGePt (top panel) in comparison with the broadened and photoionization-cross-section weighted partial DOS. The spectra were taken with 700 eV photon energy. }
\label{PESLDA700}
\end{figure}

The PES results taken at 700 eV photon energy are shown in Fig.~\ref{PESLDA700}. To facilitate the comparison with the band structure results, the experimental spectra of the LT phase (bottom panel) and the HT phase (top panel) are plotted together with their respective calculated DOS. The partial DOS are multiplied with the Fermi-Dirac distribution function, weighted by their respective tabulated photoionisation cross-sections, \cite{Yeh85} and broadened to account for the experimental resolution and lifetime effects. Finally, the commonly used integral-type of background - as indicated by the dotted lines in Fig.~\ref{PESLDA700} - is added to account for the presence of secondary electrons during the photoemission process. The cross-sections per electron at 700 eV photons are 7.4, 1.9, 3.0, and 1.7 kb/e for the Pt $4d$, Ge $4p$, Ge $4s$, and Ti $3d$, \cite{Yeh85} respectively. The Pt $5d$ and - to a lesser extent - the Ge $4s$, thus dominate at this photon energy.

A very good correspondence between the computational and the experimental results can clearly be seen in Fig.~\ref{PESLDA700}. The essential features in the experimental data are all well reproduced, including the energy gap between the Ge $4s$-like shallow core states and the remainder of the valence band. The experiment confirms that most of the Pt $5d$ spectral weight is concentrated at the high-binding-energy side of the valence band, and that the Ti $3d$ states contribute significantly to the features near the Fermi level. Most importantly, the broadening of the  Pt $5d$ and Ti $3d$ derived bands in the HT phase as compared to the LT phase is also clearly revealed by the experiment.

\begin{figure}
     \includegraphics[width=0.40\textwidth]{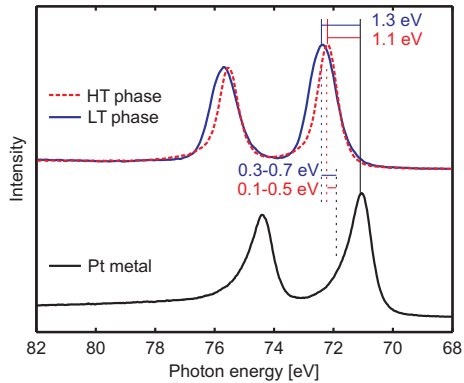}
     \vspace{-2mm} \caption{(Color online) Pt 4$f$ core level photoemission spectra of TiGePt in the low-temperature (LT, top, blue solid line) and high-temperature (HT, top, red dashed line) phase, and of elemental Pt metal (bottom, black solid line). The spectra were taken with 700 eV photon energy. Solid vertical lines represent the peak positions of the $4f_{7/2}$ levels, dashed vertical lines the center of gravity positions (see text). }
\label{CoreLevels4f}
\end{figure}

As a further check we also study the Pt 4$f$ core levels of TiGePt and compare them to those of elemental Pt. The experimental
spectra are displayed in Fig.~\ref{CoreLevels4f} and exhibit the characteristic spin-orbit splitting giving the 4$f_{5/2}$ and
4$f_{7/2}$ peaks. For the LT phase of TiGePt, the peak positions are 75.7 eV and 72.4 eV, respectively. The HT phase has peaks at 75.5 eV and 72.2 eV, while elemental Pt shows peaks at 74.4 eV and 71.1 eV, respectively. The spin-orbit splitting is thus 3.3 eV for all the three materials. This compares well with the calculated spin-orbit splitting of about 3.45--3.46 eV for TiGePt in both
modifications and elemental Pt.

In TiGePt the Pt 4$f$ peaks are shifted by 1.1--1.3 eV to higher binding energies in comparison to those of Pt metal. Similar shifts have also been observed in other noble-metal intermetallic compounds,\cite{Franco03,Gegner06,Gegner08,RosnerT} indicating a lowered averaged electron density around the noble-metal sites. To compare this chemical shift to the results of LDA calculations, one has to take into account that LDA does not incorporate many-body effects of the final state, as manifested in the asymmetric line shape in the spectra of the elemental Pt, as we will discuss below in more detail. Yet, it can be shown that final-state effects do not alter the average energy of the spectrum.\cite{Lundqvist68}  If we determine the center of gravity of the 4$f_{7/2}$, we find a binding energy of 72.4 eV for the LT phase of TiGePt, 72.2 eV for the HT phase, and 71.9$\pm$0.2\,eV for Pt metal. These centers of gravity are indicated by dashed lines in Fig.~\ref{CoreLevels4f}. Thus, the experimental chemical shift between the LT and the HT phase TiGePt, and Pt metal is about 0.1--0.5 and 0.3--0.7 eV, respectively. This is in reasonable agreement with the shift obtained from our band structure calculations which is about 0.71/0.74 eV.

We note that the line-shape of the core levels in TiGePt is not as asymmetric as for Pt metal. An asymmetry in the line-shape is caused by the presence of electron-hole pair excitations upon the creation of the core hole, i.e. screening of the core
hole by conduction-band electrons, and can be well understood in terms of the Doniac-Sunjic theory.\cite{Doniach70} The strong
asymmetry of the 4$f$ of Pt metal can therefore be taken as an indication for the high DOS with Pt character at the
$E_F$.\cite{Huefner75} The rather symmetric line shape of the 4$f$ of TiGePt, on the other hand, indicates a rather small DOS at the $E_{\mathrm{F}}$. Indeed, all this confirms the results of the valence band measurements: the main intensity of the Pt 5$d$ band is between 2 and 6 eV binding energies, with little weight at $E_{\mathrm{F}}$.

\begin{figure}
  \includegraphics[width=0.5\textwidth]{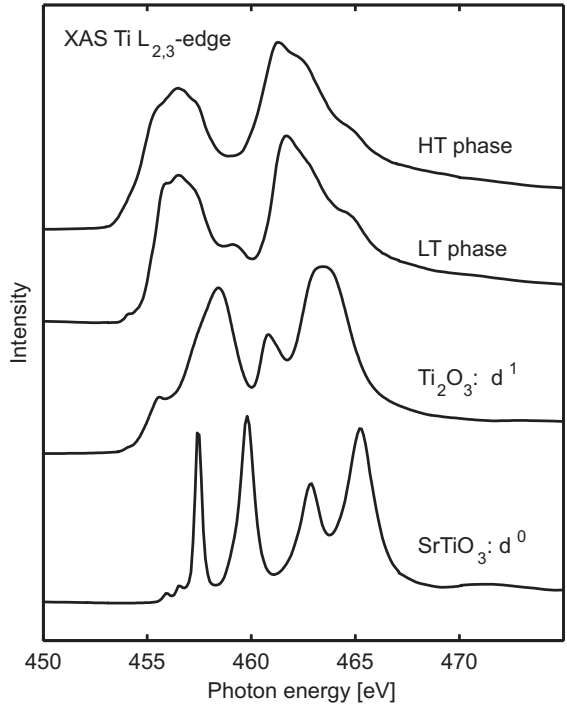}\\
  \caption{X-ray absorption spectra of the Ti $L_{2,3}$-edge of TiGePt in the low-temperature (LT) and high-temperature (HT) phase in comparison with Ti$_2$O$_3$  and SrTi$_2$O$_3$  as references for Ti $d^1$ and Ti $d^0$, respectively.}
\label{XAS}
\end{figure}

We now focus our attention to the contribution of Ti to the
electronic structure of the material. Fig.~\ref{XAS} shows the Ti
$L_{2,3}$ ($2p$$\rightarrow$$3d$) XAS spectra for the two phases
of TiGePt. It is important to note that XAS spectra are highly
sensitive to the valence state: an increase of the valence of a
transition metal ion by one causes a shift of the $L_{2,3}$ XAS
spectra by one eV or more towards higher
energies.\cite{Chen90,Hu98,Hu00,Burnus2008a} Therefore, as a
reference we include also the spectra of Ti$_2$O$_3$, a nominally
Ti$^{3+}$ ($3d^{1}$) compound, and SrTiO$_3$, a nominally
Ti$^{4+}$ ($3d^{0}$) system.

From the experimental spectra we can estimate their center of gravity, and after correcting for the background, we obtain energy
positions of roughly 460.7, 461.0, 461.5, and 462.9 eV for HT-TiGePt, LT-TiGePt, Ti$_2$O$_3$, and SrTiO$_3$, respectively.
This suggests that the valence of the Ti ions in the two modifications of TiGePt is rather similar, but appreciably
smaller than in Ti$_2$O$_3$ and SrTiO$_3$. This finding is in agreement with the effective atomic charges of titanium (+1.4 in LT and +1.3 in HT phase, respectively) obtained from the bonding analysis by means of electron density.\cite{Aackerbauer10} Further, the FPLO calculations result in the Ti $3d$ occupation of about 2.54~e and 2.53~e for HT- and LT-TiGePt, respectively, and 2.24~e and 2.14~e for Ti$_2$O$_3$ and SrTiO$_3$, respectively. These numbers follow the trend of the XAS energy positions, confirming again the consistency of the calculations.

Apart from this, one can clearly see that the spectra of Ti$_2$O$_3$ and SrTiO$_3$ show distinct multiplet structures whereas the features observed in the TiGePt spectra are much broader. This is fully consistent with the more ionic nature of the oxides as compared to the TiGePt, where covalent interactions play a significant role. In addition, the structures for HT-TiGePt are broader than for the LT phase, which is in line with our band structure calculations predicting broader bands for the metallic phase than for the semiconductor.

\begin{figure}
\includegraphics[width=0.4\textwidth,angle=0]{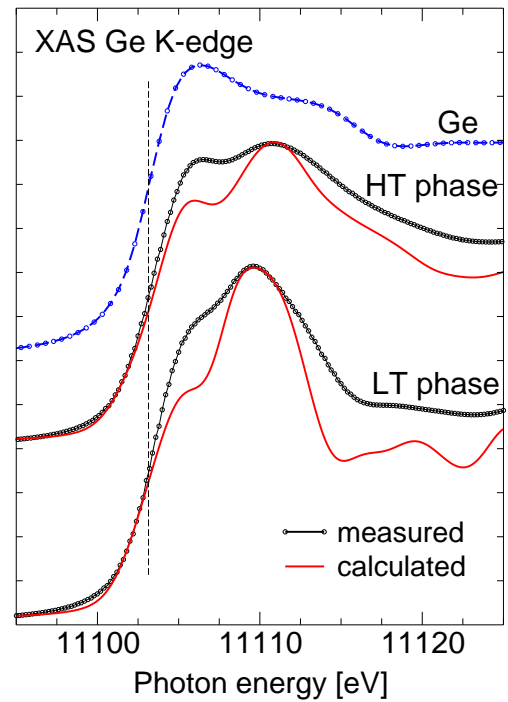}
\caption{\small (color online) X-ray absorption (XAS) spectra at the Ge $K$-edge for TiGePt in the low-temperature (LT) and high-temperature (HT) phase (black solid lines with experimental points), together with the data for the reference system Ge (blue dashed lines with experimental points) and with the calculated XAS spectra (red solid lined).
The experimental spectra were normalized using a standard method as implemented in the Athena program.\cite{Athena} The position of the absorption edges determined by taking the maximum in the first derivative of the normalized spectra is indicated by vertical dashed lines.
The theoretical curves were scaled to match the maximum in the near edge XAS region of the experimental spectra.
}
\label{XAS2p}
\end{figure}

To study in more detail the conduction band of TiGePt, we also have performed Ge $K$ near edge structure measurements. The results are displayed in Fig.~\ref{XAS2p} together with that of the elemental Ge as a reference compound.
To interpret the TiGePt spectra, we compare them with the calculated unoccupied 4$p$ partial DOS, weighted with the energy dependent transition probabilities calculated as described in the Section \ref{sec:methods}.
We can observe that most of the experimental features can be satisfactorily reproduced. The intensities are, however, not correct, but this to be expected since our calculations do not take into account the core hole effect.
It is important to note that the energy position of the Ge $K$-edge in both phases of TiGePt is the same as for elemental Ge, within the experimental error bars. This finding is in line with the basically neutral charge state of germanium in both phases of TiGePt obtained from the analysis of the electron density based on the quantum theory of atoms in molecules (QTAIM).\cite{Aackerbauer10}

\begin{figure*}
  \includegraphics[width=0.8\textwidth]{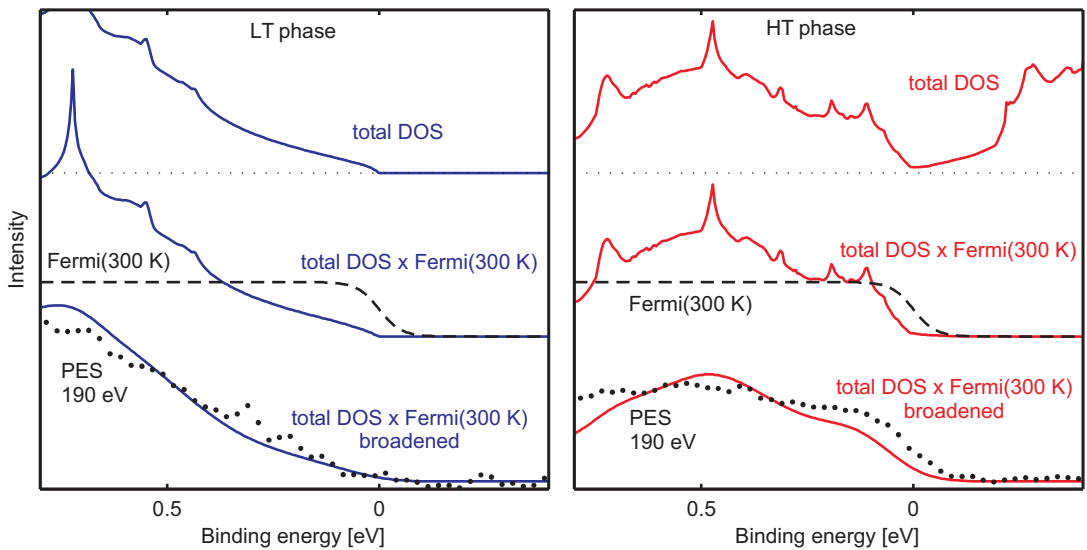}\\
  \caption{(Color online) Density of states and photoemission spectra near the Fermi level of the low-temperature (LT, left panel) and high-temperature (HT, right panel) phase of TiGePt. Top curves: density of states; middle curves: density of states   multiplied by the Fermi distribution function at 300 K; bottom curves: photoemission spectra taken using 190 eV photons (dots) and density of states multiplied by the Fermi function and broadened by the experimental resolution.}
\label{PES190}
\end{figure*}

Next, we focus on the states near the Fermi level. Fig. \ref{PES190} shows a close-up of the valence band
photoemission spectrum (dots) and the calculated DOS (solid lines) in the vicinity of the Fermi level. The photoemission spectra were taken using 190 eV photons, with an overall energy resolution of about 150 meV. To facilitate the comparison, we have multiplied the DOS with the Fermi distribution function at 300 K (dashed lines) and broadened with the experimental energy resolution.

For the LT phase, one can clearly observe a very good agreement between experiment and theory. The gentle slope and vanishing weight at the top of the valence band of this semiconductor is well reproduced. For the HT phase, the high spectral weight in the 0.2-0.8 eV region is also well explained by the theory. Yet, the observed Fermi cut-off is not in agreement with the calculated DOS. The calculations show a more reduced DOS close to the Fermi level. We currently have no explanation for this discrepancy and would like to remark that the slope of the measured spectrum in the Fermi level region matches very well the slope in the top of the calculated occupied DOS. This may suggest that the DOS of the measured material has somehow been shifted rigidly towards the Fermi level by about 80~meV. It could be that the measured material has some surface defects or imperfections which cause such a shift of the  chemical potential.

Finally, we discuss the nature of the bandgap changes in going from the LT to the HT phase.
The formation of a band gap in MgAgAs-type compounds with a valence electron count of 18 per formula unit is a well--studied issue which has been the subject of many reports within the last decade.\cite{Ogut94, Tobola96, Pierre97, Jung99, Kandpal06, Offernes06, Koehler07, Gegner08}
To get insight into the cause of the gap closure in the HT modification of TiGePt, we analyse the effect
of volume reduction first. In contrast to naive expectations, that the band broadening should decrease the gap, we find that the calculated gap size increases slightly with decreasing unit cell volume of the LT phase (10\% volume contraction leads to the increase of gap by $\sim$10\%).
Thus, the closing of the band gap in  HT-TiGePt can not be understood by solely considering the volume change.
The absence of the gap results rather from a change in Ti local environment. In LT-TiGePt, Ti atoms are tetrahedrally coordinated by Pt atoms with a short distance of 2.57 {\AA}, suggesting strong Ti-Pt interactions. Such interactions
were found to be crucial for the formation of a band gap in "half-Heusler"-type compounds.\cite{Ogut94, Tobola96, Pierre97, Jung99, Kandpal06, Offernes06, Koehler07, Gegner08}
The transition from  LT-TiGePt to the HT phase requires a breaking of the Ti-Pt bonds.\cite{Aackerbauer10}
In HT-TiGePt, the nearest neighbors of Ti are five Ge atoms with an average distance of 2.71~{\AA}, followed by six Pt atoms with a much longer average distance of 2.98 {\AA}.
The drastic change in the local coordination of Ti is reflected in the partial DOS. The sizable admixture of the Ti 3$d$ states visible for the LT phase in the binding energy region above $\sim$4 eV, resulting from the hybridization with Pt 5$d$ orbitals, is clearly reduced in the HT modification.
The essential weakening of the Ti-Pt interaction in HT-TiGePt and a corresponding increase in bonding interaction between Ti and Ge atoms have been confirmed by the combined topological analysis of the electron localizability indicator and the electron density.\cite{Aackerbauer10}

\section{summary}

We have determined the electronic structure of the low-temperature (LT) and high-temperature (HT) phases of TiGePt by means of
photoelectron spectroscopy, X-ray absorption spectroscopy and band structure calculations.
The combined theoretical and experimental study revealed substantial differences in the electronic structure for the two TiGePt modifications, although they have the same nominal composition and show similar electron counts for particular valence band states.
Most importantly, we have confirmed that the structural change in TiGePt is accompanied by an insulator-to-metal transition with an appreciable band broadening and a closing of the band gap.

The good correspondence between the computational results and the spectroscopic data for both the occupied and the unoccupied states
indicates that our calculations based on the LDA approximation provide a reasonable description of the electronic structure of the two modifications of TiGePt at ambient conditions.
Thus, the LDA level of theory can be regarded as a good starting point for a future theoretical study aiming to identify the mechanism of the structural and electronic transition in TiGePt and its driving force.

\section{Acknowledgments}

We gratefully acknowledge the NSRRC staff for providing us with
beamtime. The research in Cologne is supported by the Deutsche
Forschungsgemeinschaft through SFB 608. The authors are grateful
to Dr Ulrich Burkhardt from the MPI CPfS and Dr D. Zajac from
Hasylab for their helpful assistance during the Ge-K XAS
experiment. M.~Gam\.za is grateful for the financial support from
the DAAD foundation.

%


\end{document}